\documentclass[]{interact}
\usepackage{float}
\usepackage{epstopdf}
\usepackage[caption=false]{subfig}

\usepackage[longnamesfirst,sort]{natbib}
\bibpunct[, ]{(}{)}{;}{a}{,}{,}
\renewcommand\bibfont{\fontsize{10}{12}\selectfont}

 \usepackage{booktabs}
 \usepackage{multirow}
 \usepackage{graphicx}
 \usepackage{natbib}
 \usepackage{xcolor}
 \definecolor{blue}{RGB}{1,1,255}

\theoremstyle{plain}

\theoremstyle{definition}

\theoremstyle{remark}

\usepackage{multicol}

\begin{document}

\articletype{ARTICLE TEMPLATE}

\title{Pilgrimage to Pureland: Art, Perception and the Wutai Mural VR Reconstruction}



Pilgrimage to Pureland: Art, Perception and the Wutai Mural VR
Reconstruction

\begin{table}[h]
\centering

\begin{tabular}{p{2.5cm}p{3cm}p{3cm}p{2cm}p{2cm}}
\toprule
Name & Department & University & City & Country \\
\midrule
 \small Rongxuan Mu$^*$ & \small Architecture & \small Tianjin University & Tianjin & \small China\\\\
 \small Yuhe Nie$^*$ & \small Computer Science and Technology & \small Southern University of Science and Technology & \small Guangdong & \small China\\\\
 \small Kent Cao$^{*\dag}$ & \small Division of Arts and Humanities & \small Duke Kunshan University, Duke University & \small &  \small United States\\\\
 \small Ruoxin You & \small Arts and Humanities & \small Duke Kunshan University & \small Jiangsu & \small China \\\\
 \small Yinzong Wei & \small Chinese Literature & \small Wuhan University & \small Wuhan & \small China\\\\
 \small Xin Tong$^{\dag}$ & \small Data Science Research Center & \small Duke Kunshan University & Jiangsu & \small China\\
\bottomrule
\end{tabular}

\begin{itemize} \small
     \item [1] $*$ All authors contributed equally to this research.
     \item [2] $\dag$ Corresponding author.
\end{itemize}

\end{table}

Corresponding Author
\begin{table}[H]
\centering
\label{tab:corr}
\begin{tabular}{p{2.5cm}p{6cm}}
\toprule
Name & Email\\
\midrule
Xin Tong & xin.tong@dukekunshan.edu.cn\\\\
Kent Cao & kent.cao@dukekunshan.edu.cn\\
\bottomrule
\end{tabular}
\end{table}

\clearpage

\maketitle

\begin{abstract}
Virtual reality (VR) supports audiences to engage with cultural heritage proactively. We designed an easy-to-access and guided Pilgrimage To Pureland VR reconstruction of Dunhuang Mogao Grottoes to offer the general public an accessible and engaging way to explore the Dunhuang murals. We put forward an immersive VR reconstruction paradigm that can efficiently convert complex 2D artwork into a VR environment. We reconstructed the Mt. Wutai pilgrimage mural in Cave 61, Mogao Grottoes, Dunhuang, into an immersive VR environment and created a plot-based and interactive experience that offers users a more accessible solution to visit, understand and appreciate the complex religious, historical, and artistic value of Dunhuang murals. \textcolor{black}{Our system remarkably smoothed users' approaches to those elusive cultural heritages. Appropriate adaptation of plots and 3D VR transfer consistent with the original art style could enhance the accessibility of cultural heritages.  }

\end{abstract}


\section{Introduction}
Virtual reality creates an immersive vision with visual, auditory, and tactile sensations for participants, allowing them to observe and operate the virtual world interactively. As VR technology matures, there has been a rising trend in applying VR in the Cultural Heritage (CH) digitalization field, such as reconstructing cultural
heritage~\citep{han2019compelling,reconstructdunhuang2} and generating embodied cultural heritage knowledge~\citep{restorevr2020, relicvr2021}. 

As a real-time computer-simulated 3D environment, VR supports audiences to actively engage with CH and related art-historical knowledge rather than passively receive information~\citep{activelearning2002}. Dunhuang Mogao Grottoes is one of the world's largest and richest Buddhist art sites that survived to the present day. The murals in the caves depict a wealth of stories of Buddhist deities, secular landscapes, and religious architecture. In addition, there are significant amounts of historical documentation surrounding these murals. Therefore, it is crucial for the public to have access to the contents of these murals and understand the relevant knowledge.

Cave 61, where the Mt. Wutai mural is located, remains the largest Buddhist cave among the Mogao Grottoes~\citep{wong1993reassessment}. However, Cave 61 is currently closed to the public due to long-term conservation. As a digital medium, VR can preserve murals in various digital forms and provides an affordable and immersive solution for the public to access them. Recent studies on Dunhuang CH have been focusing on the digital reconstruction of caves~\citep{han2019compelling, reconstructdunhuang2}, allowing users to visit the caves through VR. Others focus on constructing knowledge into an embodied experience~\citep{restorevr2020}, and users can actively study such knowledge through certain gamified interaction~\citep{wutaigame2018}. Prior work suggests that VR can create a more immersive environment to support users in acquiring CH knowledge and facilitate an engaging experience~\citep{vrimmersive2015}. \citeauthor{ivr2022} reconstructed the Chinese Painting \textit{Spring Morning in the Han Palace} as a VR scene. They used Immersive Virtual Reality to refer to the virtual reconstructed environment of the artwork and verified that Immersive VR is more effective for users to appreciate, recognize and learn about CH. Most of the murals in Dunhuang Mogao Grottoes are rich in visual detail and historical stories. It is often difficult for the audience to grasp the mural well by visiting a virtual museum. An immersive VR system can provide users with an accessible solution to acquire knowledge concerning the mural. 


However, current immersive VR researches on CH still have some gaps when applied to a large-scale CH like Dunhuang. First, current immersive VR systems use a discrete narrative. Most of the knowledge is directly embedded in the immersive VR environment in the form of text rather than designing a unique VR narrative for the rich CH content~\citep{ivr2022}. This results in users being distracted by extraneous interactions in the environment and unable to focus on key knowledge acquisition. Dunhuang murals contain a wealth of historical stories and cultural information. Digital narrative in VR is an immersive exploration form to involve users in the story of CH~\citep{narrative1, narrative2}. Most of the VR experiences in CH contain a basic user flow. Some of the user flow is based on separate tasks\textcolor{black}{~\citep{restorevr2020}, which allow users to follow the guideline to finish one task or visit one place.} Others contain more intense narratives as they are designed based on relatively complete story plots~\citep{filip2022vr360storytelling}. \textcolor{black}{A carefully designed and highly interactive narrative is required to ensure that users acquire the core knowledge well. Second, many CHs contain richly detailed two-dimensional (2D) murals or reliefs, such as the Dunhuang murals. These works of art usually include a large number of complex figures, buildings, and objects and have a relative artistic style. It is difficult to completely reconstruct the 3D models of the comprehensive details of those work~\citep{reconstruction3}. Previous work usually uses manual modeling to reconstruct such 2D artwork into a 3D immersive VR environment~\citep{ivr2022, weight2021virtual, htc_2019}. These works both cost a lot of time in art modeling and have problems with fidelity. Developers have to spend a lot of time constantly calibrating and adjusting them.} Thus, an efficient way is required to convert 2D murals into 3D environments quickly. In addition, such reconstruction needs to maintain the integrity of artistic style to provide users with a highly unified and immersive environment.

To fill the gaps mentioned above, our research questions are:
\begin{enumerate}
    \item RQ1: How to design an effective narrative to improve users' acquisition of CH knowledge compared to other forms of virtual access in CH?
    \item RQ2: How to efficiently digitalize and reconstruct CH environment in the immersive VR to enhance the immersion and maintain art style?
\end{enumerate}

Our studies focus on the construction of the Mt. Wutai pilgrimage mural. We put forward a detailed paradigm for constructing an immersive VR environment through a 2D mural. Both the spectacular architecture and figures in the mural are accessible for users to acquire historical knowledge compared to other mural digitization intuitively. The narrative focuses on the Foguang Temple in Mt. Wutai, an important temple in ancient China's religious and architectural history. We compose task performance and story-telling in the VR experience, combining task-based and plot-based design. This comprehensive solution allows users to experience the historical scenes from the first-person perspective to enhance their immersive feeling and participation in the spatially reconstructed Mt. Wutai Mural. We created two scenes around Cave 61 in Dunhuang. The Cave scene is the high-definition 3D environment of Cave 61. It completely recreates a specific view of cave 61. This scene allows the user to view the 2D mural of Mt. Wutai directly. The Foguang Temple scene is the immersive VR reconstruction of the Foguang Temple. In this scene, users go through various task-based interactive stories to explore the Foguang Temple and acquire historical core knowledge. 

To verify whether the narration design and immersive VR system could address our research questions, we designed an experimental group and a control group based on the above scenes and took the Foguang Temple scene as the control variable. Through collecting experimental records, questionnaires, and interview results of 20 participants, we found that participants were aware of the legitimacy of our narrative design and the knowledge we were attempting to convey. They also understood the connection between the Foguang Temple scene and the mural and developed a strong sense of virtual embodied cognition. 
Participants in the experimental group reported significantly higher immersion levels and stated that narrative improves their cognition by familiarizing them with new knowledge. These conclusions clearly respond to our RQs.

Our main contributions are as follows:
\begin{enumerate}
    \item We put forward an immersive VR environment construction paradigm. It can provide detailed steps and encompasses efficiency, art style, and immersion. Our paradigm is easily expandable to include more immersive VR constructs of 2D CH artwork while highly preserving the details of these works.
    \item \textcolor{black}{We proposed a comprehensive narrative design that combines task-based and derivative plots} so that users can have a better sense of direct experience and focus on learning the key knowledge. These improvements will help researchers, museum personnel, professionals, and educators who wish to use VR to spread some key educational knowledge. Our work also aids the learning enthusiasm of the general public who are interested in CH.
\end{enumerate}

\section{Related Work}
\subsection{ Uniqueness and Importance of Dunhuang Grottoes and Cave 61 Murals}

 \textcolor{black}{Dunhuang art's accessibility,volume and iconography makes it the essence of ancient Chinese national culture and Eurasian culture shining on the Silk Road, presenting one of the most illustrious Buddhist art sites in the borderland of China and Central Asia~\citep{lee2012repository}. }

\textbf{Accessibility.} Even with the aid of modern transportation, traveling to Dunhuang in Northwest China is still a major decision that requires careful planning. Cave 61 is one of the “Special Caves,” which requires specific admission authorization in addition to the general Dunhuang site. The Covid pandemic further rendered travel to Dunhuang nearly impossible both at the international and domestic level in China ~\citep{wang2021history}. Therefore, high-quality digitization is urgent and crucial for access to caves and murals.

\textbf{Volume.} For the general public, the dazzling and complex detail of the Mt. Wutai Mural can be overwhelming. The Mural occupies the entire western wall of Cave 61, boasting a colossal size of 13.0 meters in length and 3.6 meters in height. A single image of such a volume is rare in global art history. Focusing on the individual motifs as well as pursuing the entire mural are both challenging tasks for the audience~\citep{xiao2008}. Consequently, struck by the difficulty of finding a comfortable point of entry, the Mural presents itself as alienating and aloof. 

\textbf{Iconography.} Hundreds of monks, pilgrims, merchants, officials, and Buddhist figures populated the Mt. Wutai Mural. Across the undulating blue-and-green landscape 195 inscriptions provide hints about the miraculous phenomena, monasteries, pagoda-stupas, placenames, Buddhist patrons, and practitioners. However, the seemingly straightforward roadmap has abounded with specialist religious, historical, and geographical knowledge. For instance, on the far-right edge of the mural, a monk stands in a square quad with a green floor, white wall, and blue roof. The explanatory epitaph is now covered in black pigment, and the information is long lost~\citep{wong1993reassessment}.

\subsection{Digital Reconstruction of Cultural Heritage}

\textcolor{black}{With computer graphic technology, the simulation and visualization to preserve cultural heritage are widely defined as virtual heritage~\citep{ibrahim2018conceptual}. The main challenge of this field is that, users are no longer impressed with basic walk-throughs. They demand more interaction, participation and content~\citep{whatnext}. Digital reconstruction of cultural heritage is a fundamental step to disseminate knowledge and value, regardless it is in the form of VR or other approaches~\citep{addison2000emerging}.} According to the differences in existing form and reconstructing strategy, tangible cultural heritages can be categorized into two types: spatial and non-spatial. The spatial heritages are in 3D forms, such as architecture and statues, which can be reconstructed with direct 3D modeling. The non-spatial heritages exist in 2D and, probably are attached to spatial ones, namely, paintings and murals, of which the digital reconstruction contains re-creating strategy such as spatialization. Therefore, the related virtual heritage projects generally cover two fields: reconstruction of spatial heritage and spatialization of non-spatial heritage.

\subsubsection{Reconstruction of Spatial Heritage}
\textcolor{black}{Most earlier virtual heritage works focused on rebuilding solid tangible cultural heritage in 3D digital immersive environment ~\citep{barreau2014virtual,flint2018virtualizing}. Many of
the examples have staged virtual heritage
offerings—beautifully textured walk-throughs of
tombs, chapels, Roman sites, and the like~\citep{whatnext}.} As for the specific case of Dunhuang Grottoes, the Dunhuang Research Academy has published a digital database that contains a selection of images of the Dunhuang caves and murals on its official website. In addition, the Academy introduced a desktop VR simulation of navigation in a handful of caves that can be experienced via PC and mobile platforms such as smartphones or tablets ~\citep{dunhuang2022}. Their works mainly focus on the documentation and representation stages of virtual heritage. Although the data is of high-quality and sufficient amount,  the approach of dissemination is neither immersive nor interactive as its VR successors. 

\textcolor{black}{\citeauthor{han2019compelling} have conducted a virtual tour of Dunhuang Cave 61, with head-mounted VR devices. Their project reconstructed the 3D interior model of Cave 61 with high-resolution texture and the exact same scale as the realistic site, which provides users with a compelling experience of the authentic conditions of the cave and murals~\citep{han2019compelling}}. Furthermore, an interactive system containing spatial context and augmented information is also implemented in the experience, offering interesting knowledge acquisition for users. In this way, visitors can have an intuitive impression of the now-missing statues in the cave and the outline of pilgrims' routes on Mt. Wutai's mural.\textcolor{black}{This work has added more interactive content in the virtual environment but is still in the basic walk-through mode.}

\textcolor{black}{As in the cases mentioned above, users engage in the virtual tour as visitors or observers, but not participants who can take part in the historical events or be brought back to the historical contexts of the mural. Thus, we aimed to bring the users back to the cultural circumstance in more immersive and embodied approaches, and attempted to break through from the simple walk-through mode in our research.}



\subsubsection{Re-creation of Non-spatial Heritage}

Fewer works cover reconstructing non-spatial cultural heritage in digital or VR contexts. Without specific spatial form, these data tend to be stored in the form of planar images or audio, such as photographs of paintings and records of folk songs. The VR representation of these works aims to transfer the planar data to the visible and spatial form.
For example, Borrowed Light Studio’s work~\citep{oculusmobilevrjam2015} \textit{The Night Cafe - An Immersive VR Tribute to Vincent van Gogh} transferred Van Gogh’s work \textit{The Night Cafe} into 3D virtual scenes, which allows users to experience the cafe in the painting spatially. The abstracted 3D rendering is the most distinguishable feature of this work. The completed work vividly represents Van Gogh’s distinctive artistic vision with the bright colors and clear brush strokes. The imagination and creation of the unpainted part of the virtual scene is another noticeable innovation of this work. To keep the scene complete, visitable, and rich in content, the author used reference material from Van Gogh and other related artists to enrich the reconstructed environment, such as rooms hidden by doors~\citep{steam}. \textcolor{black}{However, it is still only a walk-through demo, in which users can be impressed by the distinguishable art style of Van Gogh, but not much intriguing interactive content is implemented.} 
As for traditional Chinese artwork, \citeauthor{jin2020reconstructing} reconstructed the traditional painting \textit{Spring Morning in the Han Palace}~\citep{jin2020reconstructing}. Within this work, users can step into the ancient master painting to visit the splendid palace and experience seasonal activities. Fidelity was also the crux for the authors. Some pilot testees challenged the 3D modeling in the virtual scenes for the noticeable divergence of historical female figures between the VR work and painting. \textcolor{black}{The famous ancient Chinese drawing, \textit{Qingming Festival by The Riverside} is a frequently engaged material in recent works. The scale and content of \textit{Qingming Festival by The Riverside} are relatively comparable to our Dunhuang murals, which contain natural and manmade environments and various characters. The full-dome movie version~\citep{youtube_2020} of \textit{Qingming Festival by The Riverside} (displayed in the Palace Museum, Beijing ) is an innovative attempt to offer users perspectives switching from a planar drawing to a spatial virtual environment in fluent experiences. The reconstruction implemented by HTC Vive is more refined in graphics~\citep{vive_2019}, so it is more acceptable for VR headsets.  }

Therefore, in addition to reconstructing the 3D environment, our work also contains spatialization of the murals to ensure a comprehensive experience for users.

\subsection{Narrative VR Implementation in Cultural Heritage }
Most of the VR experiences implemented in cultural heritage contain a basic timeline of the user flow.  \textcolor{black}{The basic framework of the user flow is bonded with a series of tasks, which may be relatively independent in most situations. We defined this type of design mode as a "task-based" design that can be told from simple walk-throughs. Studies have shown that meaningful organization is a very powerful antecedent of learning ~\citep{regian1992virtual}. Organized with verbally meaningful plots, there are some other designs that contain more intense narratives as they are designed based on relatively complete stories. And we defined these designs as a "plot-based" mode. We referred to both the task-based design and the plot-based design to cover the educational and narrative features of CH, which will respectively work on users' knowledge acquisition and identity immersion.}
\subsubsection{Task-based VR Experience and Knowledge Acquisition}
 A basic strategy of the educational VR design is task-setting, which can improve users' engagement.

The later work~\citep{ivr2022} of the reconstruction of \textit{Spring Morning in the Han Palace} continued as an immersive VR experience to learn Chinese traditional art. The experience was designed based on a series of separated scenes with intractable components to teach user knowledge about traditional Chinese culture. The evaluation of knowledge acquisition is remarkable in this research, as they compared the head-mounted-display-based learning effectiveness with that of multi-touch tablets. 

Restore VR \citep{restorevr2020} provided a more participatory interaction for users, aimed to build a useful design system, allowing users to experience the situational restoration of murals and disseminating the specific knowledge of mural conservation. The engaging strategy decreases users’ cognitive barriers through interactive experiences. In addition, VR allows visits to the sites being conserved,  which is otherwise impossible in real life. 

Earlier work of Champion covered the effectiveness of learning in gamified VR experience. The work suggests that this solution can help decrease mistakes in navigation. But at the same time, the approach might block truly effective learning progress, especially considering the complicated mutual mechanism of learning behavior and level design~\citep{champion2006evaluating}. 

\subsubsection{Plot-based Participatory VR Narratives}

Compared to task-based VR experiences mentioned above, participatory VR narrative experiences provide more embodied first-personal participation for the users. \textcolor{black}{Interactive narrative is generally understood as approaches of digital technology to build virtual environments within which to present stories interactively, with input and feedback from the spectator. The main advantage of interactive stories is that users empathize with roles as if they were part of the
story. Adaptive stories have been taken as a way to help people connect with unfamiliar situations and characters~\citep{ostrin2018interactive}. }
\citeauthor{zhang2021meet} adapted the mythology painted in the mural of Dunhuang Cave 257, enabling the users to play as a fictional figure to protect the main character of the Deer King from completing the experience. This derivative work of the originals retains the primitive art style to strengthen its connection with the cultural heritage, simultaneously reducing users’ effort to accept the story settings~\citep{zhang2021meet}. 

Flexible adaptation also exists in the dissemination of intangible cultural heritage. The work of “Flower and the Youth” is based on VR and gesture recognition technology, with interactive performance narrative, metaphorical elements, and embodied cognition~\citep{lu2021flower}. This work provides the audience with a more immersed experience that contributes to the transmission and dissemination of traditional performing arts.

Therefore, we propose using comprehensive design strategies combining task-based design and derivative plots to conduct our research.

\section{The VR Design and Implementation}
\subsection{Design Objectives and Principles}
Our design and research objectives are twofold : (1) to create an engaging and authentic VR experience for the users and improve their understanding of the Dunhuang murals artworks; and (2) to facilitate users' awareness and knowledge of the complicated mural content through interactive narratives in VR. To achieve our goals, we followed the design principles below from a user-centered perspective when implementing Pilgrimage To Pureland environment. 

\begin{itemize}
    \item Reconstructing VR visual environments that maintain the originality of the Mt. Wutai murals in Dunhuang No. 61 Cave. Users can subsequently grasp and understand the iconography of the original murals.
    \item Creating narratives that reflect the mural content for users through interactive tasks.
\end{itemize}

\begin{figure}
\centering
\includegraphics[width=1\textwidth]{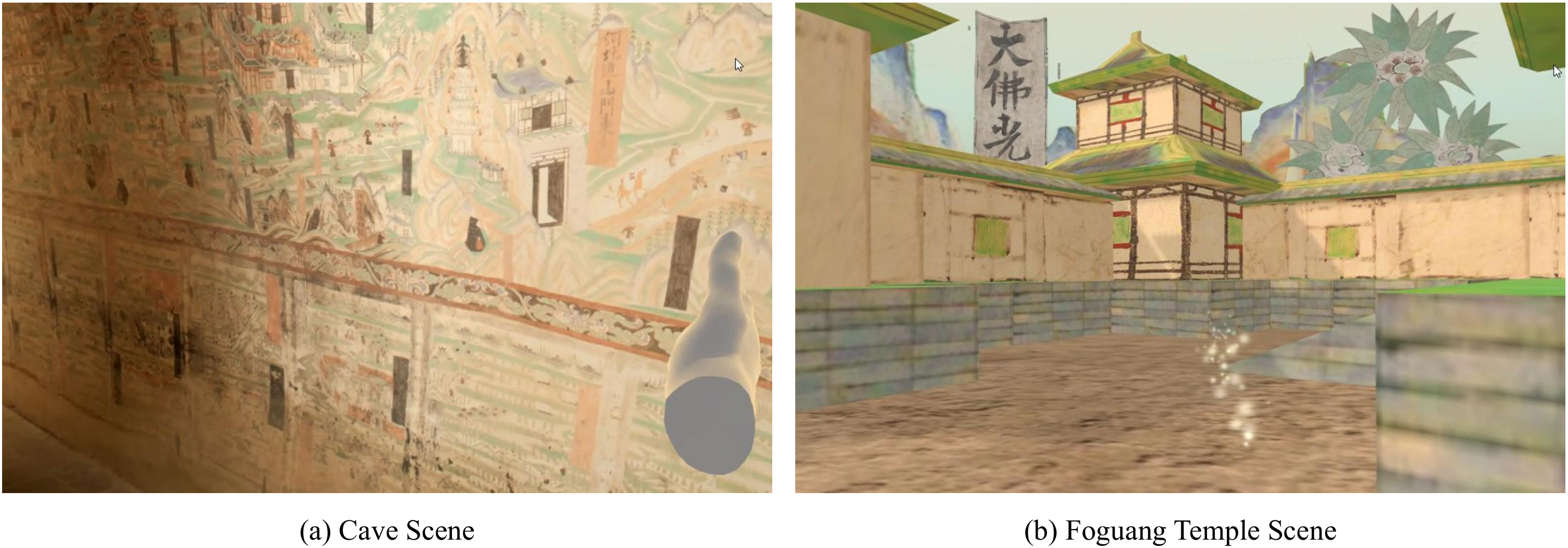}
\caption{Scene design} 
\label{SceneDesign}
\end{figure}

To improve the efficiency and fidelity of the artwork, we maintained the original 2D mural visuals from Digital Dunhuang~\citep{dunhuang2022} and reconstructed the 2D art assets into 3D objects and environments (Figure ~\ref{SceneDesign}). We referred to TOEM ~\citep{toem2021}, a manga-style adventure game developed by Nintendo, for our 2D-3D artwork reconstruction method. This game assembled 2D slices of objects and characters into 3D shapes, restricting the assembled slices to always facing the game camera. The 3D architecture fits into simple buildings built with raw geometry and rendered with a black-line shader. Therefore, we turned the 2D characters and other environmental objects (e.g., mountains and plants) into slices and resembled them into 3D objects. In addition, we further implemented the main architecture in the mural since the temples in the murals have completely symmetrical structures on all sides. See more details in Section \ref{sec:implementation}.

Mt. Wutai pilgrimage mural has a rich historical and cultural background, among which the Foguang Temple is celebrated for the discovery by Liang Sicheng and Lin Huiyin in the early twentieth century CE. Throughout history, many foreign monks once came to Wutai Mountain to worship, among which rich historical records address the Foguang Temple. We chose such a region as a blueprint to design our narration and interaction. In order to let users focus more on the core knowledge this project proposes to convey, we have integrated knowledge exposition into the task, and users need to complete the task step by step. We guided the user through conversations with characters and specific VR interaction tasks to trigger post-experience and content, giving the user layers of insight into the knowledge we were trying to transmit. In addition, to ensure that the knowledge passed on is not lost, we added a navigation system so that users can explore freely while also checking whether they have unfinished tasks.

\subsection{Core Design Features: the Scenes, Narratives, and Interactions}

\subsubsection{Scene One: The Start of the VR Experiences inside Cave No. 61}
Cave Scene is a 3D effect of real scenes, as shown in Figure \ref{SceneDesign}-a. We selected the southwest corner of Dunhuang Mogao Cave 61. The whole scene is a very realistic representation of the effect of entering the cave. Participants were able to turn their eyes 360 degrees to see the cave. Among them, they can see the complete picture of Mt. Wutai with rich details in the cave.

\begin{itemize}
    \item In the experimental group, participants will hear a speech about the historical background of Wutai Mountain. After the speech, the Foguang Temple part on the mural will have a yellow cue aperture, and when the user looks at the aperture for a period of time. They will be transmitted to the next scene.
    \item In the control group, we put the knowledge that participants needed to understand next to Mt. Wutai in the form of texts and audio. The user can turn the page by pressing a button on the handle. participants could also watch videos of the story of Liang Sicheng and Lin Huiyin. When participants think they fully understand the relevant content of Wutai Mountain and Foguang Temple, they can choose to stop the experiment.
\end{itemize}

Through this scene, we hope to enhance the user's identification as a visitor to the Dunhuang Mogao Grottoes. At the same time, users can feel the grandeur of Cave 61 and the Mt. Wutai pilgrimage mural through VR. This scenario prepares the user for the next step in the scroll. We aim to create an unrealistic artistic effect so that users have the sense of being sucked into the scroll. 

In the design of the control group, we hope that through this form, we can:

\begin{enumerate}
    \item Include the same knowledge as the experimental group.
    \item Use the same cave environment but exclude all the interactive content of the Foguang Temple
    \item Approximate the effect of users viewing web pages and videos on the traditional media platform. 
\end{enumerate}

\subsubsection{Scene Two: Developing the Narratives inside the Foguang Temple}
Foguang Temple scene, as shown in Figure \ref{SceneDesign}-b, is completely restored according to the mural of the Foguang Temple in Mt. Wutai. It is not a realistic scene. The style of the overall mural is obvious, with bright colors and a more cartoonish style. It mainly includes the Foguang Temple and the East Hall inside the temple. There are two characters to talk to and some interactive items.

\begin{figure}[H]
\centering
\includegraphics[width=1\textwidth]{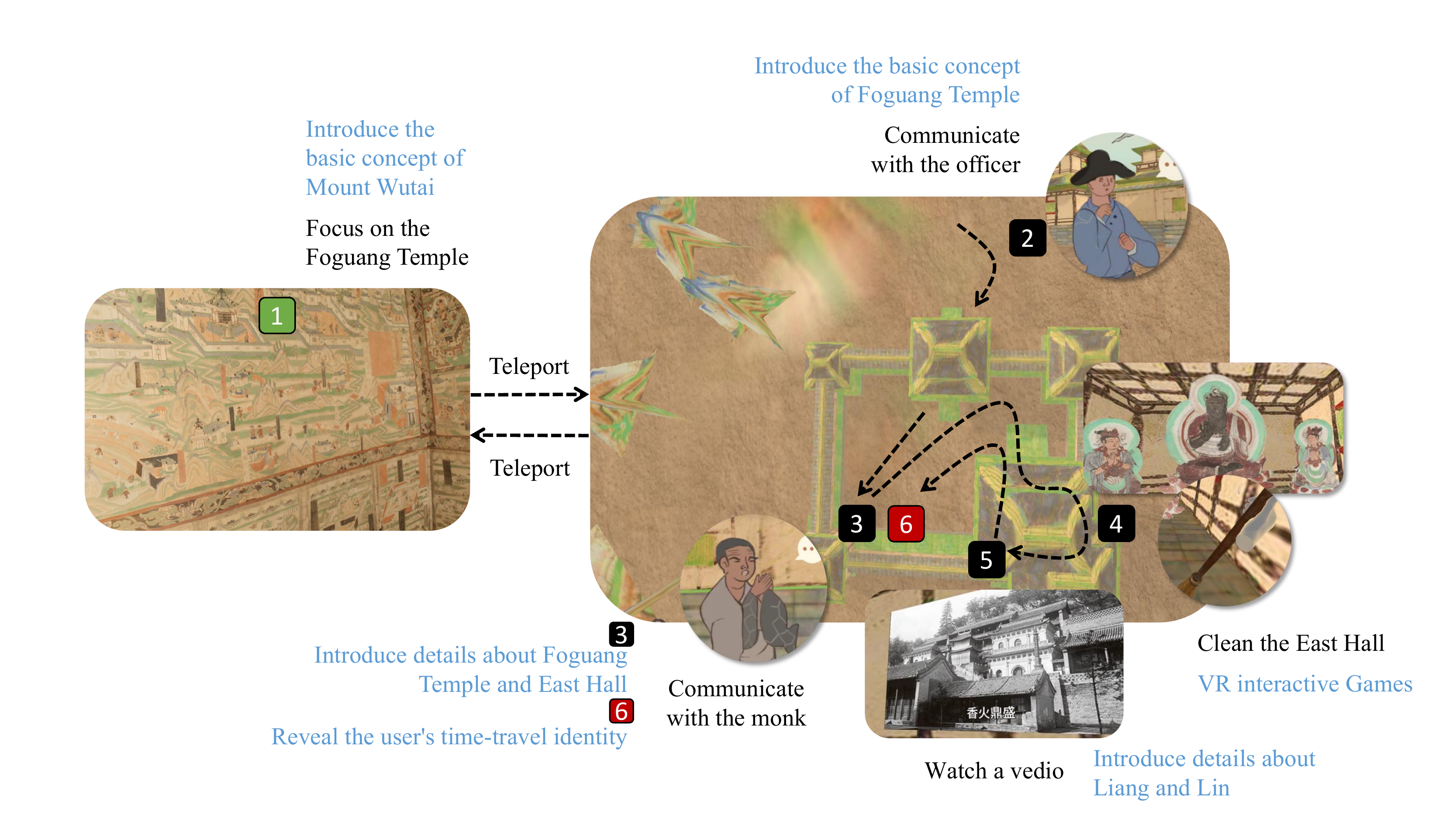}
\caption{Narrative flow diagram (1) Black text: Interaction triggers (2) Blue text: Knowledge acquisition (3) Participants start the experiment at position 1 and finish the experiment at position 6}
\label{FlowDiagram}
\end{figure}

When participants enter the Foguang Temple scene, they are free to rotate the angle of view, use the controllers to move, or move within the allowed experimental range. In addition, if participants are unsure what to do next, they can press a specific button on the handle to turn on the light-point navigation. When light-point navigation is on, a trajectory directs the participant to the next task's location.

The overall process is shown in Figure \ref{FlowDiagram}.  Participants will see an official in front of them (Figure \ref{FlowDiagram}-2). When approaching the official, participants could click on the official using the ray on the handle to interact with the figure. The official will introduce the basic knowledge of the Foguang Temple introduce the background of the temple, its chronology, historical events that occurred, etc., and invite participants to enter the temple. After entering the Foguang Temple, participants will see a monk (Figure \ref{FlowDiagram}-3) walking inside the temple. participants will be given more details about the East Hall of the Foguang Temple after talking to the monk. The monk would tell the participants that the East Hall was in disrepair and invite participants to clean it. After accepting the task, participants can enter the East Hall. In the corner of the East Hall stood a broom (Figure \ref{FlowDiagram}-4). The participants can hold the handle close to the broom, grab it, and clean the three spots where the dust fell in the East Hall. Upon completion, participants will automatically proceed to watch a documentary video (Figure \ref{FlowDiagram}-5). This video introduces the story of Liang Sicheng and Lin Huiyin, who carefully examined the architectural structure of the East Hall by cleaning the East Hall and reporting to the public the historical value of the East Hall as one of the few surviving Tang-period architectures. After watching the video, participants will walk out of the East Hall and have another conversation with the monk in the Foguang Temple (Figure \ref{FlowDiagram}-6). After a certain period, participants will be transported back to the cave to complete the whole part of the experiment.

 Through a task-driven storytelling approach, we primarily conveyed the architectural form and the history of the Foguang Temple, the story of the East Hall, Liang Sicheng, and Lin Huiyin. By communicating with NPCs in the scene, the user can obtain a basic knowledge of the Foguang Temple on the one hand and undertake the following tasks on the other hand. When the user completes the task we designed through multiple interactions, there will be new clues or NPCs to give more information. Finally, after the user experiences the main plot of the Foguang Temple, they will be transported back to Cave 61. Under this narrative background, users' avatars not only retain the knowledge regarding the Mt. Wutai mural in the Mogao Grottoes of Dunhuang but also travel back to the Foguang Temple centuries ago. Therefore, the narrative of different time periods is interspersed in the overall experience. We hope to use this scene transformation and identity cognition to enhance the user's sense of traversing time and space.

This scenario's design is to verify the immersive VR's impact on immersion and promote the user's understanding of cultural heritage history knowledge by quoting interactive narrative techniques. At the same time, we also apply immersive VR Reconstruction Paradigm to this scene.

\subsection{Technical Implementation \label{sec:implementation}}
In the construction of the immersive VR world, we hope to restore the art style, character image, atmosphere, etc., of the mural with high fidelity so that users have a feeling of being in the mural more than just visiting the Mogao Caves. In addition, we hope to reduce the cost of reconstructing 3D art resources as much as possible and reduce the difficulty of restoring the murals. In our project, there are mainly four steps during the construction of the immersive VR environment.

\begin{figure}
\centering
\includegraphics[width=1\textwidth]{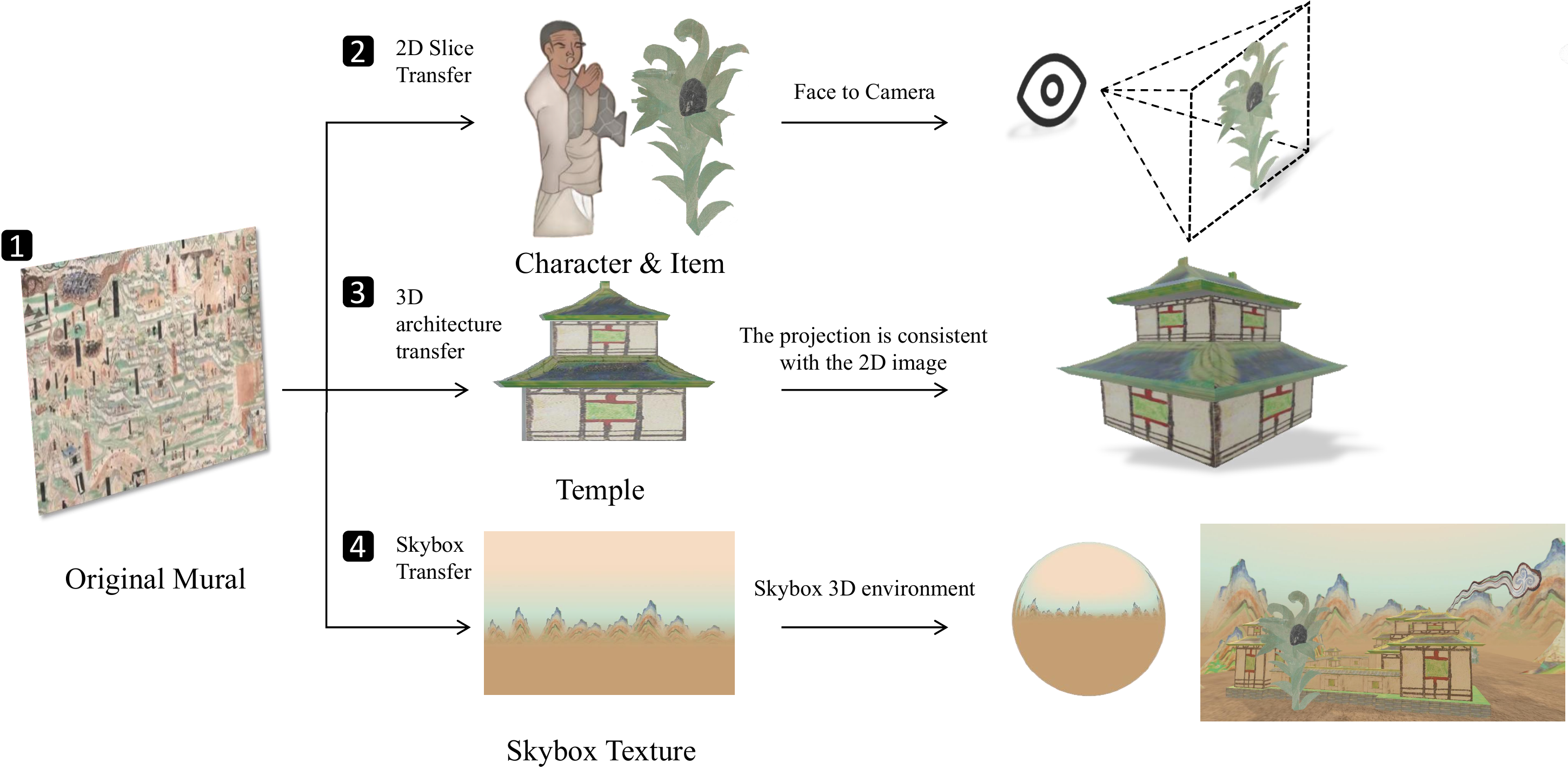}
\caption{Immersive VR reconstruction paradigm (1) Get HD images of the original mural (2) Matting 2D slices from the mural and let them face the camera (3) Matting 2D slices from the mural and construct 3D buildings that the projection of four sides are consistent with the slice (4) Post-process the mural background and integrate it into the skybox} 
\label{IVRParadigm}
\end{figure}

\textbf{Collecting Original Mural Images.} Cave 61 in Dunhuang's Mogao Grottoes is currently closed to the public to protect cultural relics. However, We can get high-definition images from the Digital Dunhuang ~\citep{dunhuang2022}. Digital Dunhuang is in the form of a digital exhibition that records up to 30 caves with more than 4,430 square meters of cave and fresco content. Its acquisition accuracy is 300DPI. Data is shared globally. We collected the 3D view cut picture of the Buddha altar north in the digital Dunhuang Cave 61 and the map-cut images of Mt. Wutai. We then concatenated these cuts to obtain a complete map of Mt. Wutai.

\textbf{The 2D slice Transfer.} We call the figures, buildings, landscapes, and other objects matted out from the original picture collectively as the 2D slice. We call the maintenance of the essential characteristics of a 2D slice and apply them to a 3D environment as a transfer. This stylized effect can not only completely retain the flat 2D characteristics of the mural, but also allow users to navigate the scene and interact with objects in 3D with minimal open cost. In a 3D environment, if a 2D slice is always perpendicular to the person's eye angle, it will give the illusion of a 3D effect. We call this form of change “face-to-eye transfer.” Based on this transfer, we can expand many dynamic effects. For example, a 2D clip keeps flipping, moving up and down, compression and expanding while maintaining face-eye transfer. 
Another type of transfer originates from a simple traditional Chinese structure of mortise and tenon joints. Two convex 2D slices are inserted together orthogonally to form a structure with a cross in the overhead view. When the objects generated by this transfer are large enough, and the user is kept in a relatively horizontal direction with these objects in the 3D environment, it will be an object of approximately the same shape when viewed from any angle. We define this transfer as a cross-transfer. In our project, all the characters and some environmental objects apply 2D slice transformation, as shown in Figure \ref{IVRParadigm}-2. When the environmental objects in the scene mainly play a decorative role, such as clouds and plants, a face-to-eye transfer can be used. A cross-transfer can be used when the environmental objects in the scene mainly play the role of obstacles, such as mountains and stones. Based on the operation mentioned above, we quickly extracted many elements from Mt. Wutai and put them into the VR environment.

\textbf{The 3D Architecture Transfer.} The main architectural structures in Foguang Temple are multi-storey temple buildings and walls, which are geometrically simple and can be reused many times. These structures should retain their 3D structural features so that the user has a clear sense of space. We built the digital models of the temple concerning both the mural and the architectural remains of the same buildings. As is shown in Figure \ref{IVRParadigm}-3, to ensure the consistency of the artistic style, we created the textures of the objects in the scene from image slices directly from the mural. And use multi-perspective observation to confirm the shape of the model matches the images on the mural from a certain perspective to guarantee the visual effect of the architectural environments.

\textbf{The Skybox Transfer.} To improve the atmosphere of the experience, we adapted the skybox of the reconstructed scene in the mural. \textcolor{black}{Skybox is a wrapper which is rendered around around the entire scene to give the impression of the horizon scenery.} The skybox texture mainly followed the Unity engine’s cube map format and was scaled and stretched into a certain ratio with the 3D module of Adobe Photoshop\footnote{Adobe Photoshop is software that is extensively used for raster image editing, graphic design and digital art.}. As shown in Figure \ref{IVRParadigm}-4, the main pattern of the horizon in the spherical skybox is mountain clips from the mural, which are duplicated and re-scaled with organic rhythm to mimic the natural environment expressed in the mural. The upper part of the texture is rendered with a light green to orange gradient color to fit in the mural's style and represent the sky's color. The lower part of the texture fits in color brown to mimic dirt and sand, which can provide a visual expansion for the earth in the mural. 

\begin{figure}
\centering
\includegraphics[width=1\textwidth]{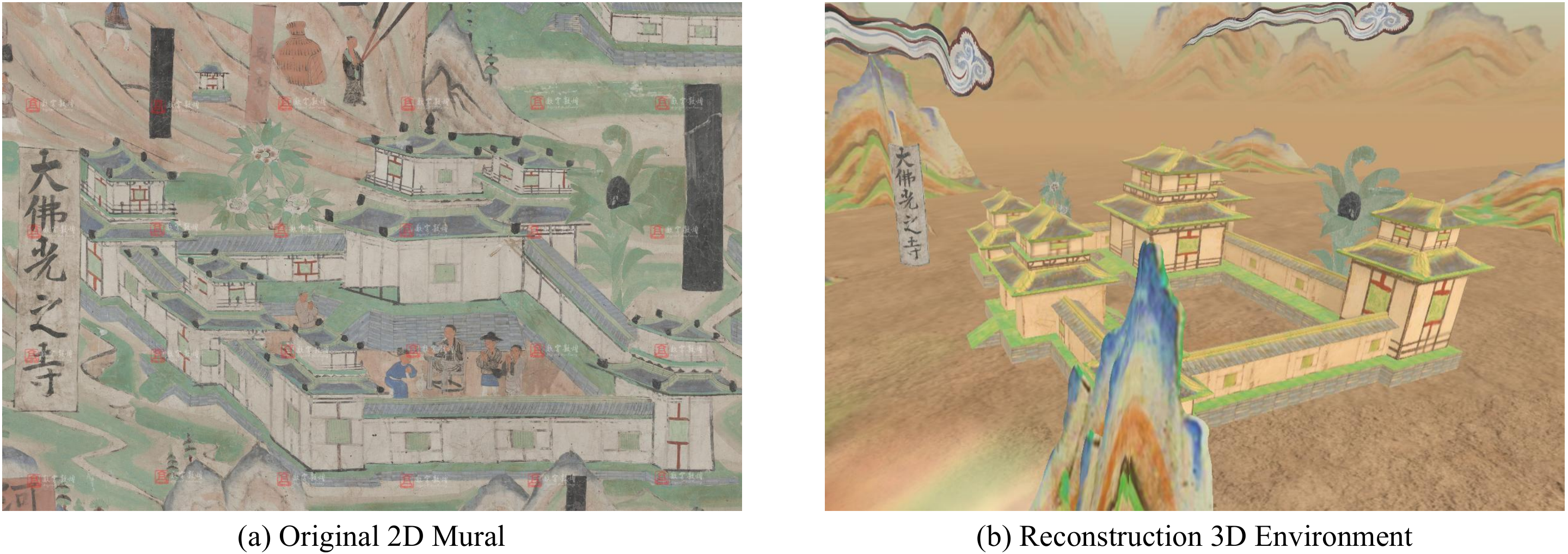}
\caption{\textcolor{black}{Reconstruction Comparison (1) The painting of the Foguang Temple from Mt. Wutai mural (2) The reconstructed immersive VR environment}}
\label{2Dto3D}
\end{figure}

\textcolor{black}{
As shown in Figure \ref{2Dto3D}, through our immersive VR reconstruction paradigm, the Pilgrimage To Pureland environment can clearly restore the original painting information and art-style of the mural.}

\subsection{Apparatus}
We use the integration toolkit Auto Hand - VR Interaction~\citep{autohand2022} in Unity3D\footnote{Unity3D is a game engine that can be used to develop video games for PC, consoles, mobile devices, and websites. We use version 2020.3.25f1c1 to develop our system.} to create the whole environment, which includes an XR system that developers can easily use to implement multiple interaction modes. And we use Blender to model some architectures. We developed our project on a computer with Intel Core i7 and NVIDIA GeForce RTX 2060. \textcolor{black}{We used stand-alone VR headsets to perform the user tests. 5 of the participants used Pico 4\footnote{Pico 4 is a VR headset developed by PICO, whose parental company is Bytedance, Ltd. } and the other 15 used Quest 2\footnote{Meta Quest 2 is also a VR headset developed by Meta Platforms.} These two devices are of the same level in related parameters like comfort, field of view and so on, therefore 
the slight differences brought by the devices could be ignored in this research. }

\section{Method}

\subsection{Participants}

We recruited 20 participants (7 male, 13 female, see Table \ref{table-participants}) through social media and workplace/university connections. The mean age of the participants was 25.5 (SD=5.92) and ranged between 19 and 40 years old. 16 of the participants reported they had experienced VR before, and 6 of them used VR headsets more than 10 times. In terms of understanding Dunhuang murals, 4 participants had been to Dunhuang and visited Cave 61 before. Of 16 participants that did not visit Dunhuang previously, 8 of them had accessed Dunhuang murals from other approaches. 19 of the participants reported they did not have any previous knowledge about Cave 61. Only one participant reported, "the cave was made to celebrate the karma of the Cao family, which was the most important family of the Guiyi military circuit," when asked about the background knowledge of Cave 61. All participants signed the consent forms and were informed that they had the right to withdraw their consent without consequences. Participants were compensated at an hourly rate of 10" RMB. 

\begin{table}[t]
\centering
\caption{Participants Demographics}
\label{table-participants}
\begin{tabular}{| c c c | c c c |}
\hline
\multicolumn{3}{| c |}{VR Narrative} & \multicolumn{3}{| c |}{Comparison system}\\
\hline
ID & Age & Sex & ID & Age & Sex\\
\hline
P1 & 37 & Female  & P11 & 23 & Female  \\
P2 & 27 & Male  & P12 & 30 & Male  \\
P3 & 22 & Male  & P13 & 28 & Male  \\
P4 & 19 & Female  & P14 & 22 & Male  \\
P5 & 21 & Female  & P15 & 25 & Female  \\
P6 & 27 & Male  & P16 & 34 & Female  \\
P7 & 20 & Female  & P17 & 40 & Male  \\
P8 & 20 & Female  & P18 & 20 & Female  \\
P9 & 24 & Female   & P19 & 24 & Male  \\
P10 & 27 & Female  & P20 & 20 & Male \\
\hline
\end{tabular}
\end{table}

\subsection{Study Design}

We designed a between-subjects experiment with two conditions: the VR narrative system and the comparison system. Our study aimed to evaluate the learning outcomes and user experience of our system. The independent variables were the VR system. We developed a traditional virtual touring system as the comparison system. The comparison system included a 360-degree panorama of Cave 61 and some interactive interfaces showing related information. In these two systems, the knowledge contained will be presented in the same order, and the content will stay identical. We hypothesized that the IVR experience would have better effects in removing the knowledge acquisition and art appreciation barrier and providing an immersive experience. The dependent variables chosen to examine the effects were measured by three aspects: 

\begin{enumerate}
    \item Accuracy of the knowledge test.
    \item Sense of presence measured SUS Presence Questionnaire.
    \item Immersion measured by the Immersive Experience Questionnaire.
\end{enumerate}

\subsection{Measurements}

To measure participants’ awareness and knowledge of Dunhuang murals, we used both quantitative and qualitative methods. The knowledge acquisition was evaluated by the score variation of a knowledge test before and after the experiment. We designed the knowledge test based on the narrative content. The test included ten questions focused on the recall and comprehension of narrative content. We gained participants' feelings and awareness through interviews. The interview questions were split into five themes: narrative, art style, VR control, understanding, and awareness of cultural heritages. 

Besides, we selected presence and immersion to examine the VR experience of our system.
We used three self-rating questionnaires to measure the presence and immersion. The presence was measured by Slater-Usoh-Steed (SUS) Presence Questionnaire~\citep{usoh2000using} and Inclusion of the Other in the Self (IOS) scale~\citep{gachter2015measuring}. The 7-item SUS presence questionnaire has been validated to evaluate presence in the virtual environment. participants rate their sense of being in the virtual environment from 1-7. The IOS scale is a single-item pictorial measure of closeness and connectedness to the participants. Participants were asked to select the picture which best described their relationship with the avatar in the VR system. We used the Immersive Experience Questionnaire (IEQ)~\citep{jennett2008measuring} to measure the immersion of our VR system. The IEQ developed by ~\citeauthor{jennett2008measuring} is a 31-item scale that measures the subjective experience of being immersed while playing a video game. Participants reported their feelings during the experience from not at all to a lot by rating 1-7.

\subsection{Procedure}

Our participants were randomly assigned to experience the VR narrative or the comparison system. \textcolor{black}{Every participant got a randomly generated three-digit number from the pre-test questionnaire. We assigned them to one of two conditions based on the parity of the generated number. To make the number of participants in the two groups equal, the last five participants who could not participate in the study in-person due to the COVID-19 pandemic were assigned to the comparison system group.}
\textcolor{black}{Participants who remotely participated in this study used their computers (with camera) and Head-mounted Displays (HMD) and were instructed to install our software. Before the experiment, researchers confirmed that the software ran successfully and the display had no difference with lab equipment. During the experiment, remote participants were guided and observed by the researchers through Zoom in real-time. While differences in the experimental environment and VR equipment may affect participants' experience, we minimized the potential effects by conforming to the VR display, instruction and experience content. Considering the narrative completely happened in the immersive VR environment, both in-person and remote participants received the same experience. As a result, the different experimental setting is unlikely to have a significant impact on participants' perceptions and attitudes towards the experience.}

\subsubsection{Consent Form and Pre-test}
Before the experiment, the participants were first provided with a consent form. By signing the consent form, the participants agreed to record video, audio, and images of their participation in the experiment. The participants then completed the pre-test questionnaire, including demographic questions and the knowledge test, after providing consent. Data was collected through the Qualtrics online questionnaire system. Participants were then guided to put on the headset and start the VR experience.

\subsubsection{VR Experience}

\begin{figure}
\centering
\includegraphics[width=1\textwidth]{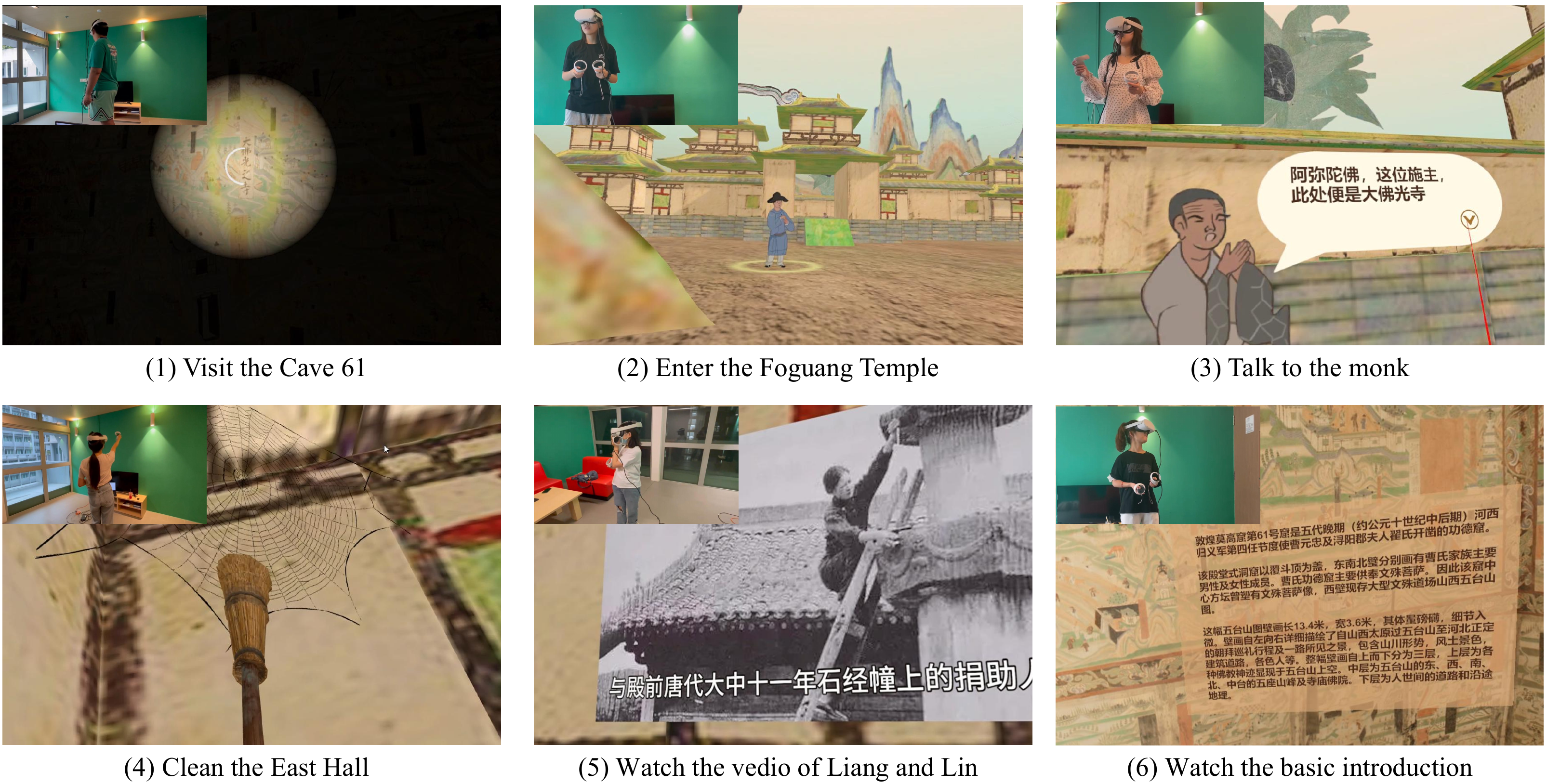}
\caption{participants experience Pilgrimage To Pureland (1-5) VR narrative system (6) Comparison system} 
\label{experiment}
\end{figure}

As shown in Figure~\ref{experiment}, researchers would briefly introduce the control methods of VR controllers and the system's interactions. After participants confirmed their full understanding of VR control, they were guided to enter the system and start their experience. Participants spent about 15 minutes experiencing the VR narrative system or the comparison system. In the VR narrative system, participants followed task-based storytelling and explored the Foguang Temple of Mt. Wutai. Participants were asked to finish all the tasks in the VR narrative. While in the comparison system, participants entered the virtual Cave 61 as a visitor. Participants listened to an audio tour and followed the light spot, which highlighted the part of the mural related to the audio content. The virtual tour in the comparison system ended as the playing of the audio introduction finished. We carried out the study by watching the participants on the monitor and recording the whole process for later analysis. \textcolor{black}{For remote participants, we asked them to open their computer front camera and share screens of HMD in Zoom. We observed their experience and recorded the whole process using Zoom's built-in recording function.} 

\subsubsection{Post-test}
After the experience, the participants completed the post-test survey on the Qualtrics online questionnaire system. The survey included \textcolor{black}{four} parts: the knowledge test, SUS Presence Questionnaire, IOS, and IEQ. In the end, we conducted semi-structured interviews with participants. The procedure lasted about one hour.

\subsection{Data analysis}

We used mixed methods to collect and analyze the experimental data. We collected participants’ raw accuracy scores (out of ten) of the knowledge test before and after the experiment. We used descriptive analysis and the one-way repeated-measures ANOVA to analyze the interaction effect of TIME * GROUP. We also collected participants' responses to the SUS presence questionnaire, IEQ, and IOS. We used the descriptive analysis and independent sample T-test to compare the difference between the two conditions. 

For the qualitative data collected from the interview, we used inductive thematic coding~\citep{fereday2006demonstrating} to identify potential principals. Two researchers coded the data independently in the first round. The primary codes include usability, narratives, awareness and knowledge, and art style. We then went back to the data and conducted one round of focused coding to reach a consensus on the themes. All the quotes used in the paper were translated from Mandarin to English by two authors and checked by the other authors.

\section{Results}
\subsection{Knowledge Test}

\begin{figure}[H]
\centering
\includegraphics[width=0.6\textwidth]{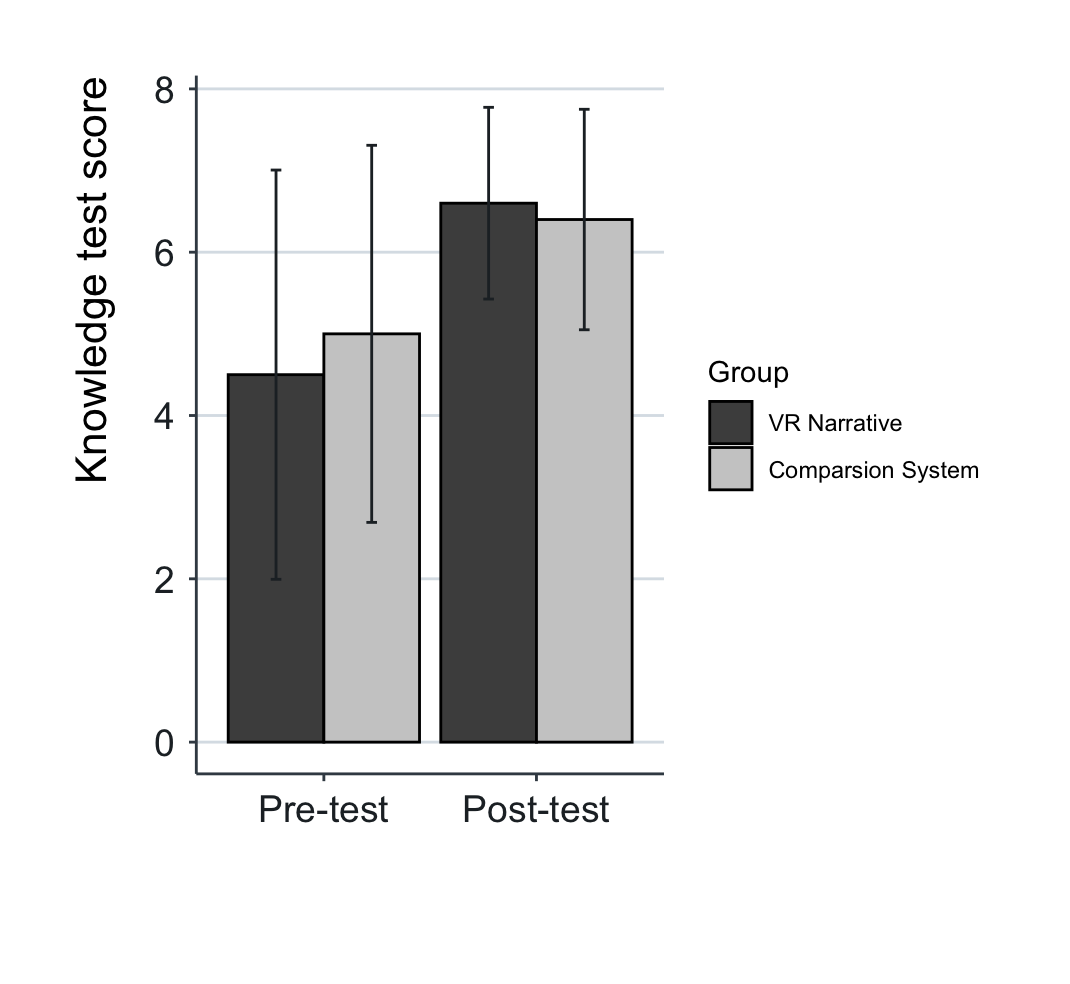}
\caption{Mean knowledge test score in pre-test and post-test of each condition. Error bars  indicate standard deviations.}
\label{KnowledgeTest}
\end{figure}

We conducted a one-way repeated measure to analyze the interaction of time (pre-test and post-test) and conditions (VR narrative and comparison system). \textcolor{black}{It can be seen in Figure \ref{KnowledgeTest} that participants got the higher score on the knowledge test after (VR narrative: M = 6.60, SD = 1.17; Comparison system: M = 6.40, SD = 1.35) than before the experiment (VR narrative: M = 4.50, SD = 2.51; Comparison system: M = 5.00, SD = 2.31), F(1,18) = 22.36, p $<$ 0.001.} These statistical results showed the effort of the VR experience on knowledge acquisition. However, the effect of time on conditions had no statistical significance (F(1,18) = 0.89, n.s.). One potential explanation revealed from the interview data is that the VR narrative system involved more details in the experience design. The increasing information brought by the abundant experiential content might increase the cognition cost when acquiring knowledge. More details as discussed in section 5.3.2.

\subsection{Self-reported Ratings}
\begin{figure}[H]
\centering
\includegraphics[width=1\textwidth]{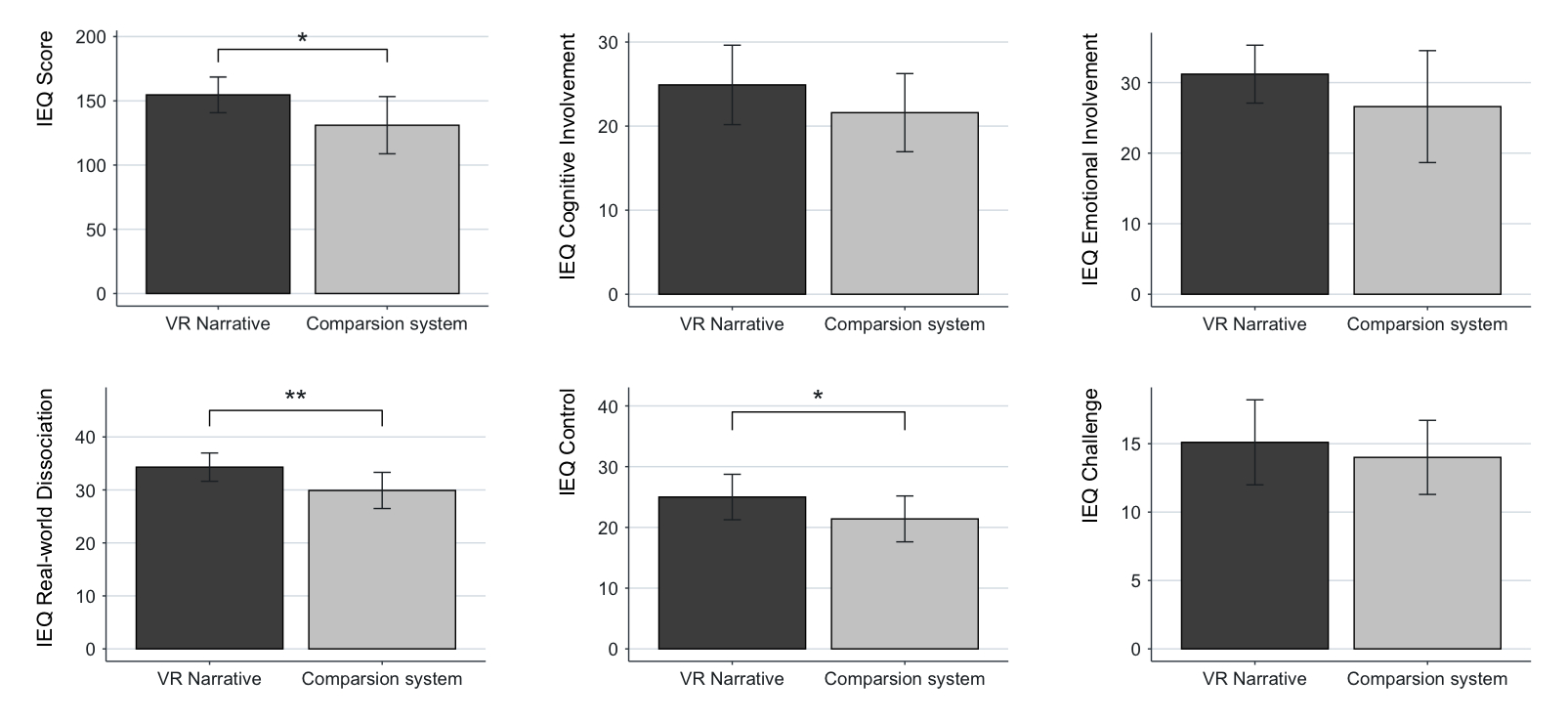}
\caption{Mean IEQ scores in each condition. Error bars  indicate standard deviations.}
\label{IEQ}
\end{figure}

We then compared the questionnaire result by independent sample T-test.  No significant difference was shown in the SUS score (t(18) = 1.04, p =.31) and IOS (t(18) = 0.91, p = .38) in the two conditions. This may be because participants’ limited VR experience forced them to pay extra attention to control the virtual avatar during the experience. Among 20 participants, 4 participants had never experienced VR headsets before, and 9 participants had very limited experience with VR headsets (less than 5 times). The unfamiliar interaction methods of VR headsets might distract them from feeling present in the VR environment. Besides, two participants reported negative feelings towards the virtual hands and the bloom task in the interview.\textcolor{black}{ The unrealistic interaction might also have a negative effect on presence. In terms of immersion, participants reported a more immersive experience in the VR narrative system than in the comparison system, which can be seen in \ref{IEQ score}.} The results showed that our VR narrative (M = 154.60, SD = 13.90) had a better performance than the comparison system (M = 131.00, SD = 22.25). The difference was statistically significant (t(18) = 2.85, p = .01)(\ref{IEQ}). The results of IEQ sub-factors suggested that there was a significant difference in control (t(18) = 2.14, p = .05) and real-world dissociation (t(18) = 3.21, p $<$ 0.001). The results indicated participants felt more controllable and were less aware of the real-world surroundings ~\cite{jennett2008measuring} during the VR narrative experience. There is no statistically significant difference in the other three sub-factors, including cognitive involvement (t(18) = 1.57, p = .13), emotional involvement(t(18) = 1.63, p = .12), and challenge(t(18) = 0.84, p = .41). One potential reason is that our narrative was based on Buddhist stories, which proceed in a peaceful way and was less familiar for the participants. Therefore, the participants might feel less challenged and cognitively and emotionally involved during the experience.

\begin{table}
\centering
\caption{IEQ scores in each condition.}
\begin{tabular}{@{}lrrrrrl@{}}
\label{IEQ score}
\toprule 
\multirow{2}{*}{IEQ} &
  \multicolumn{2}{c}{Mean} &
  \multicolumn{2}{c}{SD} &
  \multicolumn{2}{c}{T-Test} \\ \cmidrule(l){2-7} 
 &
  \multicolumn{1}{c}{VR} &
  \multicolumn{1}{c}{Com} &
  \multicolumn{1}{c}{VR} &
  \multicolumn{1}{c}{Com} &
  \multicolumn{1}{c}{t} &
  \multicolumn{1}{c}{p} \\ \cmidrule(r){1-7}
Immersion               & 154.60 & 131.00 & 13.90 & 22.25 & 2.85 & 0.01*   \\
Cognitive Involvement   & 24.90  & 21.60  & 4.72  & 4.65  & 1.57 & 0.13    \\
Emotional Involvement   & 31.20  & 26.60  & 4.10  & 7.93  & 1.63 & 0.12    \\
Real-world Dissociation & 34.30  & 29.90  & 2.67  & 3.41  & 3.21 & 0.005** \\
Control                 & 25.00  & 21.40  & 3.74  & 3.78  & 2.14 & 0.05*   \\
Challenge               & 15.10  & 14.00  & 3.11  & 2.71  & 0.84 & 0.41    \\ \bottomrule
\end{tabular}
\end{table}

\subsection{Qualitative Feedback}
\begin{figure}[H]
\centering
\includegraphics[width=0.8\textwidth]{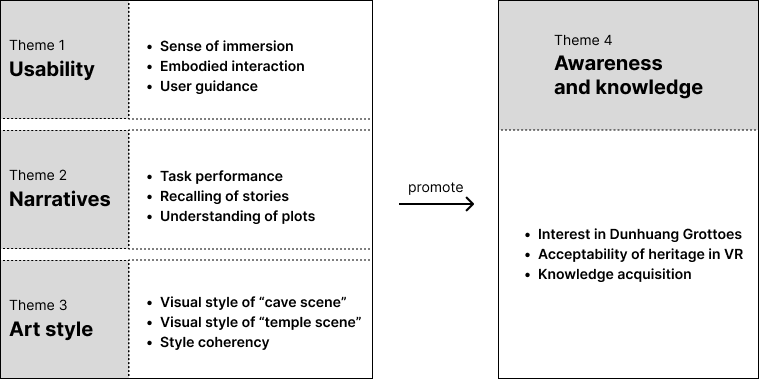}
\caption{Diagram of Themes and Subthemes} 
\label{Theme Diagram}
\end{figure}

With inclusive collection and analysis of their words and ideas, we work out the participants’ qualitative feedback, which mainly covers these four themes: usability, narratives, awareness and knowledge, and art style (Figure.\ref{Theme Diagram}).
\subsubsection{Usability grows with Immersion, Interaction, and Guidance}
Usability influences the users’ general experience, which is one of the most complicated aspects to address in this project. As we added more content and details, more difficulties and barriers in usage could be brought about. We roughly divide the feedback from the participants into items about immersion, interaction, and guidance to analyze in depth.

\textbf{Immersion can be Improved with the Details of Content.} The sense of immersion decides the basic quality of a VR system as we aim to bring the users to a different scene from their realistic surroundings, and it is largely dependent on the details of its content.
\textcolor{black}{In our interview, most participants (N = 7/10) in the VR Narrative group agreed that their experiences in VR were strongly immersive, but only half (N = 5/10) think so when it comes to the comparison system group.} This corresponds to the quantitative analysis in section 4.5, although it is difficult to judge whether the narrative system itself improved the users' feeling of immersion from such limited data samples, considering other factors such as the time length and users’ individual differences. Still, we can acquire some meaningful information that helps to research immersion from the participants’ words.
 
The general feeling of realism of VR is beyond most of the traditional media techs like printed media and electronic screens but still is far from beyond the true reality, at least with today’s technology level. The quote from participant ID P11 is representative of most of the participants’ ideas:
\begin{quotation}
“\textcolor{black}{……because there are still some limitations of technology, there is the feeling of a 3D movie or a 4D movie, nonetheless it is no way to completely make you feel like you're really standing there. The lighting is definitely much better than the real cave.}”
\begin{flushright}——Participant ID P11\end{flushright}
\end{quotation}

They admit the gap is caused by technical limitations between VR and the real world. At the same time, some unexpected benefits brought by VR are also acknowledged. 

The completeness and exquisiteness of the scene also influence users’ feelings of immersion. More details, whether realistic or stylish, can help to attract users’ attention and drive them to interact with the environment rather than pause and think about the authenticity of the scene or  even cause some discomfort. Some participants complained about the quality of the details in the scene:
\begin{quotation}
“……The main thing is the \textcolor{black}{graphics. The current VR graphic quality is not very good. } It’s rough and doesn’t feel very immersive. I might choose 4 (in 10 points to evaluate the immersion)……maybe because it’s short-lived. If it’s a game, it’s a strong plot, and the sense of substitution\textcolor{black}{(substitution of the participant's identity with the 
virtual characters)} will be stronger, even if it’s not a VR……”
\begin{flushright}——Participant ID P2\end{flushright}

\end{quotation}

The participants who gave high scores about the immersion also take the artistic quality as a reference. They might just be more compatible and tolerant with the conditions of visual effect, especially considering the relatively refined texture of the skybox in the cave scene, which was taken with professional panoramic devices and keeps high fidelity of the real site. From Participant ID P2’s quote above, the plot or task setting may also influence the immersion, but it varies with the audiences’ personal tastes. Different users might expect different stories, which is difficult to have a solution to unify and satisfy everyone’s needs.
\begin{quotation}
“……added some interactive activities such as cleaning, and then there are many different places that need to be cleaned, on the ground, and then there are spider webs on the ceiling. These are very rich activities, so it feels more immersive.”

\begin{flushright}——Participant ID P3\end{flushright}
\end{quotation}

As mentioned in Participant ID P3’s quote above, the influence taken by the design of interaction and tasks is also fundamental to users’ experience of immersion and other aspects. We will address it in detail in the next subsection. 

Other factors, such as motion sickness and discomfort caused by the weight of devices, were also mentioned by several of the interviewees (ID P2, ID P7, ID P9),  but these factors are somehow dependent on the users’ individual differences, here we will not go deep with them.

\textbf{Interaction Balances Learning Efficiency and Fluency of Experience.}
\textcolor{black}{Most of the participants (N = 18/20) in both groups felt
easy with the interaction system, but because of the more comprehensive tasks with continuous plots for the VR Narrative group to perform, they had more negative feedback about usability in our interview.} Some of the disadvantages could bring some challenges to the participants but would not block their information acquisition. 

As in some specific tasks, the mechanical and physical effects of the interaction could disappoint the users. The interaction of the virtual hand and the broom might be too strong to handle for some of the participants(like ID P5 and ID P2) because they felt it is not realistic enough to control, and they may also dislike the task setting itself. But some other participants (ID P1 and ID P3) favor the current task and interaction settings with the broom.

\begin{quotation}
“And what I don't like the most is that the old monk asked me to use a broom to clean the east hall of the Foguang Temple. \textcolor{black}{I don't know if it's because of my operation or what, it's a bit stuck in movement, I'm not sure about that. It is not very comfortable to operate.}”
\begin{flushright}——Participant ID P5\end{flushright}
\end{quotation}

\begin{quotation}
"I like cleaning the most. Cleaned up. Although I don't like it very much in real life, I think it is a more interactive place. Yes, so it makes me feel that to be in VR is to complete a task, and I think it is quite happy……Because I feel like when I was just cleaning, for example, I had a feeling that I would pick up a broom when I was cleaning at home. And when I look around, I actually feel that it is similar to the real scene, and it will not be dizzy."
\begin{flushright}——Participant ID P1\end{flushright}

\end{quotation}

And admittedly, the interaction details are still far from smooth and need to be polished in future iterations:

\begin{quotation}
"......the broom that was used to sweep the large items might be(hard) because of some obstructions between the models. For some people who have no experience in VR, this may not be particularly convenient. There are spider webs on the ceiling. This is very innovative, but it's a little hard to spot."
\begin{flushright}——Participant ID P3\end{flushright}
\end{quotation}

Also, the trigger and pause mechanism of the dialogue system with NPCs also caused some more attempts to acquire enough information to finish the plots.

\begin{quotation}
"......when I meet the character NPC inside and then when I'm talking, if I accidentally touch the move button, I move away from him a little bit, then the dialogue will end, and I have to go to him again to restart that conversation. It's just that it's a little uncomfortable."
\begin{flushright}——Participant ID P5\end{flushright}
\end{quotation}

\begin{quotation}
"......and then see if you can break two or three sentences and combine them into one long sentence. Right. But I also think that actually short sentences are more like real conversations. That's just to say if it's a bit......(redundant)in terms of operation."
\begin{flushright}——Participant ID P1\end{flushright}
\end{quotation}
However, as designers, our design to make the dialogue as complete and progressive steps to finish the conversation was out of the aim that the users can have enough time to understand and memorize the meaningful words as educational information in the brakes of each sentence of the NPCs. Here also represents the trade-offs between experience and efficiency in our design, hoping to convey the knowledge in an intriguing way.

Other aspects of the interaction design were also mentioned by our participants, like the  locomotion system in different scenes, the feedback in different senses, and controller key settings; these items are rather general in this project. 

\textbf{Guidance Fits the Main Task.}
\textcolor{black}{The guidance for users of the two groups was set to be different, as the comparison system group contained very simple interaction, so basic visual guidance helped the participants throughout the whole experience, while the VR narrative group with narrative plots was much more complicated, and so was its guidance. Combined with dialogue introduction, visual marks, and path guiding, exquisite as the guidance system is, users are still confused with some of the operations.} 

\begin{quotation}
"Yes, but it is because the background is orange-yellow, and then the light has bright yellow, but it is not particularly bright, so I am actually in the middle. I just discovered it and saw it, which means that he is not so eye-catching."
\begin{flushright}——Participant ID P16\end{flushright}

\end{quotation}

\begin{quotation}
"In fact, if I was facing the front, I could see it. But I might not have noticed if I had my back to that place. But she has a loop all the time. It doesn't end all at once. I think it may be better to combine a little sound effect.
…… I think the color might be very similar to that panoramic 360-degree scene that its overall reality sees. So I feel that the color is actually not so easy to notice. Maybe it will be more obvious if you use another strong contrast search."
\begin{flushright}——Participant ID P15\end{flushright}
\end{quotation}

As mentioned in the quotations above, the color of the visual mark is difficult to tell from the background of the scene, and the audio guidance seems insufficient in the cave scene. Similar problems exist in the reconstructed temple scene but did not block users’ experience due to our real-time help. Complicated but not systematically organized guidance design in both the VR narrative and comparison system groups is a remarkable limitation of this project at an early stage to be solved in future work.

Although some of the functions might not be the most user-friendly solution for specific crowds, there are plenty of trade-offs in the details of the experience design, and we need to keep a balance between usability and our initial design aims. Admittedly, there are still some severe design flaws in the experience that might influence the core experience and need to be solved with more iterations of work, like insufficient guidance within the design and some accidental malfunctions in the programs. However, these temporary risks and drawbacks come with abundant experiential content. It is not the choice of the direction of narrative design that causes these faults, and it is just because the more we create, the more we need to fix them. This can partially explain the unremarkable relation between the knowledge test results and the two different types of VR experience because problems grow with the system we build. As a more intriguing way we designed to teach, the way itself may cause an increase in cognition cost.

\subsubsection{Narratives Improve Cognition by Familiarizing New Knowledge}
Narrative design smooths users’ experience of learning about new knowledge. All participants can remember the rough basic structure of the timeline of the VR experience in both the VR narrative and comparison system group, which respectively takes around 10 and 4 minutes to finish. 

\textcolor{black}{However, in detail, the VR narrative group  participants (N = 10) tend to have more effective information entries for feedback in our interview than the comparison system group.} Basically, it is because the archeological and religious content is elusive for most of the participants, as when asked about the confusing experiences, some participants pointed out the historical and religious knowledge was plenty and difficult for them:
\begin{quotation}
“ The content is a bit difficult for me because maybe I'm not that Buddhist, and so, maybe a little bit lacking in historical knowledge.”
\begin{flushright}——Participant ID P1 from the VR narrative group\end{flushright}

\end{quotation}

\begin{quotation}
“ What I find more confusing is that it seems that a large part of the wall is talking about the content that they worship because I don't know much about this aspect. Then, if I read it carefully, in fact, if it is not combined with his explanation, that is, if I read it by myself, I can't really understand what story he is telling.”
\begin{flushright}——Participant ID P15 from the comparison system group\end{flushright}

\end{quotation}

Relatively familiar content only appears in the interactive experience within the narrative storyline in the VR narrative group, from which the users could organize more words to describe the subtle actions they performed and the variable scenes they saw. 

Both two sets of experiences improved the participants’ cognition of the cultural heritage of the Dunhuang Mogao Cave 61 mural and some other related cultural concepts. Most of the participants showed their interest in introducing the mural and other cultural heritage to be represented in VR. They appreciated the application of VR technology and were willing to experience different cultural heritage via VR. Although there is no remarkable difference in the effect of culture population within the two sets of experiments, the feedback of some participants can tell us some subtle distinctions between the two scenes (realistic cave scene and reconstructed temple scene) within the VR narrative group of narrative experience.

\begin{quotation}
“……Art style, cave or? It is very majestic, but in the temple, it feels very friendly as a whole, and then the affinity is very simple……”
\begin{flushright}——Participant ID P3\end{flushright}  
\end{quotation}

\begin{quotation}
“……the game is generally made a little more cartoony, that is, when you enter the scene of the Foguang Temple, there is actually a little bit of such a feeling, that is, there will be a little abstract expression, and then people feel that it is easier and clearer. That is, he may have the advantage of being a more cartoonish image, which will make people more intimate, and will not feel that it is very difficult to understand.”
\begin{flushright}——Participant ID P5\end{flushright}
\end{quotation}

These two quotations are selected from the answers to the question related to the art style. From the two participants’ feedback, we can figure out that the first-personal narrative scene is somehow more friendly and with more affinity to the users. And the second quote (from participant ID P5) takes it as a gamification and argues it may somehow decrease the cognition threshold of the audience. Also, within the reconstruction, the fixed visual perspective of the mural was changed, which can be confirmed as the source of the feeling of affinity; a brand new view from the pilgrim was created from the original drawings. Therefore the audience can get close and more engaged in the mural, but not in an unsatisfying vulgar way, which is undeniably a significant improvement for the cultural population.

\textcolor{black}{Therefore, the VR narrative group participants tend to be more impressed by the VR experience than the comparison system group by the content of the VR experience. Even 2 participants in the comparison system group (ID P14 and ID P16) could not list any item that impressed them in the cave as we first asked them for their favorite elements in the VR experience, which is partial because of the simple and straightforward display of information without intriguing plot design.} In this way, narratives can reduce cognitive barriers for users and smooth the experience in new environments.
\subsubsection{Art Style Integrity Strengthens Completeness of Experience}
Most of the participants (N = 19/20) were able to recognize the art style in the Cave 61 reconstruction scene as realistic and antique, which is attributed to the real-site captured panoramic pictures. This technique conserves the original visual effect of the murals. The details of the murals were completely represented to the participants, so even if they did not have much knowledge about the cultural heritage, the colors and lines on the drawings could still touch the audience. One of the participants in the comparison system group (ID P13) even directly named out the exact dynasty of the artwork, saying 
\begin{quotation}
“……another particularly impressive one is a Buddha statue on my left at that time, and his entire artistic style was that of the Northern Wei Dynasty. Basically, the Northern Wei style was dominated by lines.” 
\begin{flushright}——Participant ID P13\end{flushright}
\end{quotation}

The dynasty he named, Northern Wei, is exactly the age when Buddhism flourished in the region of the Wutai Mountains for a time in history; it also matches the main content of the Wutai Mountains murals in Cave 61.  

As for the art style in the scene of Foguang Temple, which is reconstructed by our creative team based on the transition of 2D images on the mural into cubic models and circumstances, most of the participants in the VR narrative could understand the artistic  connection between the recreated scenes and the original mural of the architectures and environments. Basically, the connection in the participants’ cognition is built with the similar colors of the two scenes. As participant ID P1 said,

\begin{quotation}
“......I think the relationship is quite integrated. That is, the one in his mural is actually the color used last time and the actual one. Indeed, I think it can be quite integrated.”
\begin{flushright}——Participant ID 234\end{flushright}
\end{quotation}

Only one participant (ID P8) that did not realize the connection also agreed with the integrity of the two parts in the experience as a whole, answering

\begin{quotation}
“Is it related? ......I forgot what the narrator said at the time. But I wouldn't find it awkward.”
\begin{flushright}——Participant ID P8\end{flushright}
\end{quotation}

Besides, as mentioned in section 5.3.2, the style of the Foguang Temple scene is recognized as cartoonish by several participants (ID P3, ID P5), which seems quite different from the description of the cave’s style. With further analysis, we found that it is caused by the 2D slice usage in the 3D scene and the bright and abstract color of the environment we created. These factors do not mean the reconstructed circumstance is divergent from its original source. It is exactly the mural’s style  itself. The former “realistic” comments described the spatial feeling of the caves but not the murals themselves.

However, due to limited technology and time, the details of both the panoramic cave scene and the reconstructed  temple scene were complained about by the participants. The most remarkable complaint is the resolution of our textures of the models is not high enough, which might increase the users’ discomfort and reduce their feeling of immersion. Also, the relatively humble animations of NPCs in the VR narrative group were named as a drawback.
Still, the integrity in the style of the reconstructed scenes and the original mural in the cave powerfully shows the effectiveness of our design in keeping the essence of the cave’s art. Especially the participant’s positive feedback on the color consistency is also straight proof of the validity of our design techniques to use original digital clips of the murals as the texture of the reconstructed environment in the 2D-3D transfer. This consistency in art improves users’ feeling of the VR experience as a whole.

\section{Discussion}
Our research is designed to reconstruct an immersive VR experience to enhance the public's knowledge about cultural heritage with a general methodology that constructs 3D scenes from traditional 2D painting images. The design of the narrative experience and technical methods of transferring are the main focuses of our project and will be discussed in depth below.

\subsection{Narrative Experience Reduces Cognitive Barrier}

\textcolor{black}{In response to RQ1, we chose to use narrative design to substantiate the otherwise elusive historical knowledge to users.} This innovation effectively reduced cognitive barriers and increased the sense of immersion for users. Although the rearranged derivative story plot could be more coherent and intriguing, it  earned remarkable positive feedback from the experiment participants. The intuitive game-like experience lets users get close to the unfamiliar but world-famous art. However, due to the individual difference in learning behavior, the effect of specific knowledge acquisition is not evidently improved with the narrative system.

Some related works also provide users with task-based VR experiences to teach users artistic knowledge or techniques in cultural heritage conservation~\citep{ivr2022,restorevr2020}. These works took separated sequences in a whole scene, which is similar to what we conducted in our control group: tasks or visual-audio information in a certain order, but never organically arranged in continuous plots. Other reconstruction work may take the original plots into the VR experience, which are based on the root works. That is to say, the paintings themselves contain story-telling plots, and the VR scenes they built work as stages for the stories to unfold.

Our work re-composed the raw materials from the murals and created the first-person script for users. \textcolor{black}{The concept of "interactive narrative" combines narratives with games and interactivity with non-linearity. We normally think of a narrative as a series of events that are linked causally and chronologically, the flexibility of space-time coordinates can also be taken as a potential feature in the VR context\citep{elsaesser2014pushing}.} The advantage of this approach lies in the reduced dependence on the original plot, which may not exist in some non-narrative original references. This game-like secondary creation strategy with more familiarity and engagement for users makes our project distinct from the earlier studies.

\subsection{Faithful 3D Interpretation to the Original}

\textcolor{black}{To cover RQ2, we addressed a series of issues related to 3D presentation.} The reconstructed VR environment's integrity to its original references can help users to build a complete cognition of the cultural heritage's essences, which is fundamental for educational purposes. In our research, the integrity of the derivative design and its origin is basically represented through the uniform art style. By clipping the mural's images to make textures for the 3D models, we realized the remarkable consistency of the visual effect. According to observation and interviews, the users highly praised the consistent color and shapes of the architecture in murals and our reconstructed VR scenes. The key contribution of the uniform art style is to provide users with smooth, continuous, and coherent visual experiences.
 
Related work also strengthens the authenticity of derivative work to the original source. A faithful reconstruction of the spatial environment is relatively easy to achieve. However, authenticity becomes a great challenge when it comes to animated characters such as figures. First, the requirement for the reconstruction of figures is more demanding. The users tend to be more perceptible to figures than the wider environments, as figures are usually more interactive and close to users, especially some in human form. If a tree or a building is not sufficiently authentic as the origin, most non-expert users will pay less attention to the differences. However, if a human figure appears less naturalistic, it is much easier for users to point out visual awkwardness. Second, the given information about the figures is far from sufficient. Generally, in traditional planar paintings, artists can only present one specific view of the figures, resulting in the reference being too limited to build a convincing 3D model. Furthermore, all the figures in the original paintings and murals are stationary. It is difficult for designers to imagine and create their actions and movements with our own experience. 
 
Pilot participants of \citeauthor{jin2020reconstructing} 's study argue the early reconstructed figures resemble Disney cartoons, which provide a sprightly experience but are not implicit and tranquil as the original painting. These deviations could bring about inconsistency in users' cognition and experience. To avoid this situation, our strategy is to keep the figure reconstruction minimally intrusive to the original. We directly transformed the characters from 2D planar clips in the virtual scenes and kept them constantly facing the users' cameras to ensure that the figures looked identical to the ones in the murals. The feedback is also positive, as our reconstructed figures are more vivid and intriguing than static murals, keeping the coherency in art with the original, given the planar animations are still not naturalistic enough. Our 2D-3D mixed transferring solution for VR reconstruction can be taken as an effective approach in the later work.

\subsection{Efficient Reconstruction Strategy Aids Large-Scale Scene Production}
\textcolor{black}{
The VR reconstruction of cultural heritage mostly adopts 3D modeling. The reconstruction of real scenes can usually be done by manual modeling or existing scanning techniques. As for pure 2D and large-scale scene reconstruction, such as murals, manual modeling can only be adopted. In large-scale scenes, manual modeling is time-consuming and costly and requires more consideration in terms of authenticity and performance problems. Our reconstruction strategy can effectively solve these problems.}

\textcolor{black}{
Our strategy is devoted to capturing the same original slice (2D clip) from the mural, which relieves the modeling burden and addresses authenticity issues. In addition, the strategy can be well integrated with existing AI technology. Using the computer vision semantic segmentation technology, we can implement the program generation of these 2D clips, 3D architectural structure maps, and Skyboxes and further reduce the art costs. Moreover, since our strategy uses 2D clips and 3D models built with simple geometry as much as possible (based on mural content), our reconstruction can almost ignore the performance overhead of 3D models and generate mural scenes on a large scale. It is not required to have models with different LoDs in order to optimize the performance, which provides a better solution for the reconstruction of a complex and delicate cultural heritage. Our paradigm can be extended to other cultural relics of mural painting, such as Yongle Temple, St. Lawrence Church, and San Bartolo. Since all the 2D paintings, especially the mural, can be extracted and transformed through our paradigm.
}

\subsection{Limitations and Future Work}

The limitations of this research are as follows:
First, the experimental samples are relatively limited, both in the aspects of experience content and the number of test participants.
Second, the Guidance design is relatively insufficient in the entire system, which increases the difficulty for users. Third, the artwork in VR scenes and GUI is not delicate enough. \textcolor{black}{As the resolution of textures is not high enough, some details of the virtual environment turned out to be rough and vague.}

Prospective work in narrative VR experience to improve the cognition of cultural heritage about Dunhuang Cave 61 Murals requires us to increase the experiment samples with a larger number of participants and with a more polished design. We plan to improve our design in the following directions: First, enrich the plot of the narrative VR experience to cover more content of the original murals. Second, polish the guidance system to help the new-hand user get familiar with the VR operation system. Third, improve the quality of artwork in the VR scenes, namely, improve the design of GUI, enhance textures' resolution, and optimize the characters' animation. Fourth, figure out more general solutions for effective knowledge acquisition in VR environments.

\section{Conclusion}

\textcolor{black}{The conclusion of this research mainly covers two aspects: 1) the value and effect of narrative design in VR systems, and 2) the digital reconstruction and translation of traditional artworks. Although knowledge acquisition via VR is a relatively complex and uncertain process to evaluate, the improvement in awareness and understanding of cultural heritage is remarkable through our VR experience design.} Narrative design with adapted plots in VR experience can effectively improve the acceptability of cultural heritage, for example, the Dunhuang murals in Cave 61, to non-specialist audiences. With adapted plots and gamified first-personal experience, users will become familiar with advanced-level knowledge of history, religion, and art in an improved approach. The transferring of original planar references to spatial VR scenes can be effectively fulfilled with a mixed-methods solution: reconstruct environmental elements into 3D models with high-fidelity to their original colors and shapes, and make more interactive characters 2D animated clips as they appear in the original sources. Excessive alteration and beautification are not expected in the reconstruction of traditional planar art forms. It is supposed to be faithful to the original.


\bibliography{reference}
\bibliographystyle{apalike}

\end{document}